\documentclass[11pt]{article}

\usepackage[english]{babel}

\usepackage[a4paper,top=2cm,bottom=2cm,left=3cm,right=3cm,marginparwidth=1.75cm]{geometry}

\usepackage{amsmath}
\usepackage{amssymb}
\usepackage{graphicx}

\usepackage[colorlinks=true, allcolors=blue]{hyperref}
\providecommand{\doi}[1]{\href{https://doi.org/#1}{https://doi.org/#1}}

\usepackage{multicol}
\usepackage{authblk}
\usepackage{enumitem}

\usepackage{lmodern}

\usepackage{titlesec}
\titleformat{\paragraph}[block]{\normalfont\normalsize\bfseries}{\theparagraph}{1em}{}
\titlespacing*{\paragraph}{0pt}{3.25ex plus 1ex minus .2ex}{1.5ex plus .2ex}
\usepackage{setspace}
\usepackage{fancyhdr}
\usepackage[numbers,sort&compress]{natbib}
\makeatletter
\newcommand{\TOCtitleonly}{%
  \section*{\contentsname}%
  \thispagestyle{fancy}%
}
\newcommand{\TOCbodyonly}{%
  \@starttoc{toc}%
}
\makeatother

\title{Reflective Empiricism: \\Bias Reflection and Introspection as a Scientific Method}

\author{Oliver Marc Wittwer}
\affil{Independent Researcher, Weesen, Switzerland\\
	\small{Formerly ETH Zurich (Dr.\ sc.\ nat.)}\\
	ORCID: \href{https://orcid.org/0009-0003-2205-1159}{0009-0003-2205-1159}\\
	Corresponding author:
	\href{mailto:oliver.wittwer@gmail.com}{oliver.wittwer@gmail.com}}

\date{June 6, 2026 \\ \small{Version 2.0 (preprint)}
}

\begin{document}

\bibliographystyle{unsrt}

\maketitle
\thispagestyle{fancy}

\renewcommand{\thefootnote}{}
\footnotetext{Version 2.0 (preprint), drafted April 2025, updated and uploaded here June 2026. Version~2.0 extends the original with a new Section~\ref{sec:outcomes}, presenting case studies of the author's subsequent works that demonstrate the method's interdisciplinary scope, and updates the outlook accordingly.}
\renewcommand{\thefootnote}{\arabic{footnote}}

\vspace{-1em}
\begin{abstract}
This paper introduces \textit{Reflective Empiricism}, an extension of empirical science that incorporates subjective perception and consciousness processes as equally valid sources of knowledge. It views reality as an interplay of subjective experience and objective laws, comprehensible only through systematic introspection, bias reflection, and premise-based logical-explorative modeling. This approach overcomes paradigmatic blindness arising from unreflected subjective filters in established paradigms, promoting an adaptable science. Innovations include a method for bias recognition, premise-based models grounded in observed phenomena to unlock new conceptual spaces, and Eureka moments---intuitive insights---as starting points for hypotheses, subsequently tested empirically. The author's self-observation, such as analyzing belief formation, demonstrates its application and transformative power. Rooted in philosophical and scientific-historical references (e.g., Archimedes' intuition, quantum observer effect), \textit{Reflective Empiricism} connects physics, psychology, and philosophy, enhancing interdisciplinary synthesis and accelerating knowledge creation by leveraging anomalies and subjective depth. It does not seek to replace empirical research but to enrich it, enabling a more holistic approach to phenomena that have not yet been fully grasped. A subsequent body of the author's work is presented as case studies demonstrating the application of the method introduced here.
\end{abstract}


\clearpage
\TOCtitleonly
\begingroup
\setstretch{1.1}
\TOCbodyonly
\endgroup
\clearpage

\section{Introduction: The Need for Reflection}

Modern science owes its remarkable successes to the empirical method, which prioritizes objective, reproducible results. However, this focus on pure objectivity is increasingly reaching its limits, particularly with phenomena inseparably linked to subjective experience---such as consciousness, unexpected anomalies, and boundary phenomena \cite{chalmers1995facing}. These phenomena often elude purely objective observation and are easily marginalized or dismissed as irrelevant in established paradigms, or established theories are adhered to even when contradictory evidence is present. This sometimes leads to heated, biased debates that distract from the actual arguments and underlying facts, even though these objectively suggest an expansion of existing theories.

\subsection{Limitations of Traditional Empirical Methodology}

The history of physics shows that scientific progress is often driven by paradigm shifts, each of which has transcended the limits of previous thinking. Yet, despite these advances, modern science remains largely confined to an objectivist worldview rooted in Cartesian dualism. While this focus on objective methods has elucidated the material world with impressive precision, it increasingly reveals weaknesses when it comes to phenomena that cannot be grasped solely through external observation.

\subsection{The Illusion of Pure Objectivity and the Problem of Unreflected Bias}

This paper represents a bold and necessary challenge as it questions not only individual scientific assumptions but also the broader mindset of researchers, aiming to address blind spots and expand scientific inquiry. The author argues that the empirical method reaches its limits because it marginalizes subjective experiences and prematurely dismisses anomalies as ``anecdotal''.

Empirical science strives for a supposedly pure objectivity, which, however, remains an illusion: Every observation, every hypothesis, and every result is filtered through the subjective lens of the researcher, shaped by their worldview and beliefs \cite{feyerabend1975against}. This unnoticed bias distorts reality instead of revealing it. As a historical example, one might note the geocentric view of Aristotle, which distorted the observation of the heavens \cite{omodeo2016cosmology}.

\subsection{A Plea for Reflective Empiricism: Expansion and Complement of Science}

The plea for \textit{Reflective Empiricism} aims to close this gap by expanding and making science adaptable. It recalls the intuitive genius of Archimedes or Bohr's epistemological contributions to quantum physics \cite{vitruvius1960ten,holton1973roots}. \textit{Reflective Empiricism} is not just a methodological innovation but a return to the intuitive roots of groundbreaking discoveries in science. It invites us to overcome the limitations of a restricted objectivism and to develop a more comprehensive perspective that allows not only psychological but also physical and philosophical phenomena to appear in a new light.

\section{Author's Background: Interdisciplinary Synthesis}

As a physicist with a diploma and a doctorate in natural sciences (ETH Zurich), the author brings the necessary toolkit to investigate phenomena through precise observation and analytical thinking. Simultaneously, he draws on 29 years of experience exploring the interface of consciousness and physics. His intensive interdisciplinary engagement with topics in biology, psychology, philosophy, technological developments, and artificial intelligence enriches his expertise and allows him to integrate diverse perspectives.

This work represents one of the central outcomes of his interdisciplinary pursuit to create a synthesis in science between cognition, psychology, depth psychology, mathematics, physics, and spirituality.

Through his practice of intensive self-observation---often in states of deep, reflective introspection where thinking, feeling, and perceiving merge---and analytical reflection that incorporates external phenomena, \textit{Reflective Empiricism} emerged. It is not merely a method but a bridge that transcends the boundaries of the traditional scientific paradigm and seeks to grasp reality as a whole. \textit{Reflective Empiricism} is reminiscent of the philosophical reflections of Leibniz and the groundbreaking approaches of Einstein, both of whom uniquely bridged disciplines in their pioneering work.

\section{Blind Spots of Empiricism}

Empirical science strives for a supposedly pure objectivity, which, however, remains an illusion---a legacy of Cartesian dualism that continues to shape modern research \cite{chalmers1996conscious}. Researchers who consider themselves objective are often guided by subjective blind spots---be it through deeply rooted paradigms, institutional constraints such as publication pressure, or the tendency to dismiss unusual observations as unscientific. This distortion manifests itself in a systematic practice: The burden of proof is subtly shifted to those who propose new ideas or anomalies, while established science often hesitates to investigate these phenomena---a pattern deeply rooted since Bacon's inductivism. As in the metaphor of ``The Emperor's New Clothes'', an obvious truth is ignored because it challenges the prevailing worldview \cite{kuhn1962structure}. Historical examples, such as the discovery of penicillin by Alexander Fleming, show that true innovation often arises from the courageous exploration of the unexplained, not from the mere confirmation of the known.

While historically necessary, the attempt to understand the world purely objectively-–-an inheritance of the 17th century and the Enlightenment---fails today due to the inherent distortions of the subjective lens, a problem that empirical science often overlooks, thereby limiting its own scope. Historically, the emphasis on objectivity was essential to overcome the dogmatic constraints of scholasticism. However, this one-sidedness has led to the marginalization of phenomena such as consciousness. Only through the integration of both aspects---subjective perception and objective laws---can a complete understanding of reality be achieved.

\section{Resistance to Reflective Ways of Thinking in the History of Science}

The history of science is rich in examples of resistance to new ways of thinking. Galileo's heliocentrism was condemned by the church because it threatened the geocentric worldview \cite{sharratt1994galileo}; Alfred Wegener's continental drift theory was initially ridiculed because it did not fit into the static models of geology of his time \cite{oreskes1999rejection}. Historical examples like Galileo and Wegener illustrate how resistance to paradigm shifts has delayed progress, despite the eventual acceptance of their revolutionary ideas. In both cases, contemporaries perceived the questioning of their worldview as an attack on their established beliefs---a reflex that is human, but which reveals precisely the weakness criticized by the author.

Similarly, the intuitive insight of Archimedes, who recognized the principles of hydrostatics in a bathtub, or Einstein's mental leaps to the theory of relativity, show that great breakthroughs often stem from subjective moments, later validated empirically. Historically, breakthroughs such as Newton's gravitation or Einstein's theory of relativity emerged in moments when researchers set aside their intellect and allowed their intuition to guide them in search of truth and understanding. This highlights that \textit{Reflective Empiricism} embodies the essence of scientific creativity by systematically integrating intuition and empirical validation \cite{westfall1980never,isaacson2007einstein}.

This resistance not only delays the recognition of potential breakthroughs but also hinders the development of new, more comprehensive theories. This demonstrates the necessity of a mindset that systematically integrates such intuitions, as \textit{Reflective Empiricism} aims to do.

\section{Reflective Thinking: Overcoming Bias and Expanding Knowledge}

\textit{Reflective Empiricism} is not a rejection of empiricism, but an extension that acknowledges subjective experience and consciousness processes as legitimate sources of knowledge. It redefines the responsibility of researchers by calling on science to actively investigate anomalies and boundary phenomena---whether the nature of consciousness or recognized phenomena like spontaneous remission---instead of shifting the burden of proof solely to those who propose new ideas.

This practice of placing responsibility entirely on challengers of existing paradigms hinders progress. Reflective Thinking, by contrast, encourages courageous exploration of the unexplained, as Fleming demonstrated with penicillin. Recognizing and dissolving the researcher's own perceptual filters enables science to develop new models of reality and fosters interdisciplinary openness between previously incompatible fields of knowledge.

To remain capable of learning, science must not only strive for reproducibility but also systematically investigate patterns that challenge current understanding. Only in this way can we expand the boundaries of knowledge and grasp reality in its full complexity.

\subsection{Core Principles of Reflective Empiricism}

\textit{Reflective Science} complements and expands empirical methodology by recognizing the subjective dimension of the researcher as an indispensable part of knowledge acquisition. The following describes the core principles in comparison to the empirical method, with a look at their position in the history of science:

\subsubsection{Reflective Observation (Introspection)}

In contrast to empirical science, which focuses exclusively on external data collection through experiments, \textit{Reflective Science} additionally directs the view inward. It explores the consciousness processes of the researcher and uses reflective introspection as an additional source of knowledge. A reflective introspection is essential to consider patterns and regularities of belief patterns as an integral part of the knowledge acquisition process.

This method recognizes that every observation is colored by subjective perception, which in turn is shaped by individual beliefs and experiences---a principle that early natural philosophers like Descartes intuitively understood but was suppressed in modern science.

\subsubsection{Premise-Based Model Building}

\textit{Reflective Empiricism} uses premise-based logical-explorative modeling as a central tool. Instead of relying exclusively on empirical data, speculative assumptions (premises) based on subjective insights or observed phenomena are allowed and used as a starting point for the development of coherent explanatory models. These models are not primarily bound to falsifiability but serve to explore new conceptual spaces and generate new hypotheses (see Section \ref{sec:modellbildung}).

\subsubsection{Subjective and Intersubjective Validation}

While empirical methodology relies on external validation through experiments, \textit{Reflective Empiricism} initially uses subjective comprehensibility as a criterion. Insights are gained through inner evidence and intuitive understanding and subsequently transferred into an intersubjectively comprehensible framework (similar to mathematics). The final validation ideally occurs through empirical verification of the hypotheses and experiments derived from the models.

\subsubsection{Holistic Thinking (Integration of Logic, Intuition, Emotion)}

\textit{Reflective Empiricism} overcomes the limitation to purely logical-rational thinking. It integrates intuition, emotions, and bodily sensations as equally valid sources of knowledge (``feeling thinking and thinking feeling''). This holistic approach makes it possible to grasp the complex interactions between consciousness, perception, and reality. Feeling thinking and thinking feeling refers to the conscious integration of intuition, emotions, and bodily sensations into the scientific thinking process, in order to enable a more comprehensive understanding of the phenomena to be investigated.

\subsection{Eureka Moments: Catalysts of Knowledge}

Eureka moments---moments of immediate, intuitive insight---are central catalysts in \textit{Reflective Empiricism}. They arise when the researcher, through continuous self-reflection, penetrates his subjective filters, which allows him a direct, holistic grasp of the underlying dynamics. These moments are not only an expression of intuition, but also a starting point for the systematic investigation and formalization of insights. They serve as anchor points to transfer subjective insights into an intersubjectively comprehensible framework. In contrast to the empirical method, which primarily relies on external confirmation, \textit{Reflective Empiricism} is rooted in the inner evidence of such Eureka moments.

This approach is similar to the understanding of mathematical relationships, in which true comprehension goes beyond mechanical reproduction \cite{polanyi1958personal}, or the intuitive insights of Archimedes, who recognized the principles of hydrostatics in a moment of bathing. Such moments arise when the researcher, through reflection, penetrates the subjective filters of his perception and grasps the objective laws in interaction. Their frequency among knowledge carriers could be a measure of the robustness of a finding, as future work on subjective cognition could show---a perspective that could expand the history of science to include the role of intuition.

While such insights can often be formalized afterwards, a formalization before the intuitive grasp of the inner dynamics is hardly possible.

\section{Introspection: Reflecting on Bias}

\subsection{Example: Reflective Insight into the Formation of Beliefs}

The author developed a method of conscious introspection, enabling him to regularly move to a meta-level of consciousness, a nodal point where thinking, perceiving, observing, and feeling converge. As one example among many insights gained by the author through Reflective Research, a chain of observations on the emergence of beliefs and behavioral patterns shall be summarized here.

In this state, the author imagined the perception of an infant, experiencing the world much like a child in a cinema for the first time and being completely absorbed by the film. For the infant, everything merges---colors, shapes, and feelings exist as a unified whole, with no reality outside these impressions. Every moment is so captivating that attention flows freely, without forming a distinct will---a state of pure presence.

Over time, the will awakens: The infant seeks to drink, touch, and explore, but encounters resistance---for example, when the mother acts against his wishes. This creates tension: ``I want to, but I cannot''. Initially, the stress is suppressed, but soon reactions such as rebellion, withdrawal, or anger emerge, leading to beliefs like, ``If I cry loud enough, I get what I want''. Layer by layer, such patterns shape identity---some harmless, others destructive.

Later, in experiences like first love, these patterns surface. The pain caused to others mirrors unresolved childhood pain. This creates a conflict: Should one suppress this empathetic awareness and continue the behavior, or adjust one's needs to avoid causing harm? Each decision in such interactions shapes the character, becoming an expression of past, often repressed choices. These decisions are not erased; in emotionally triggering moments, they resurface with full force, causing us to react like the hurt or defiant children we once were, despite being adults.

Through self-reflection, these patterns can be uncovered and transformed. They often appear as ``energy games''---struggles for power, validation, or control. With this awareness, the world appears like a multiplayer game, where intentions, decisions, and actions create dynamic interactions with others. The reactions of the outside world are revealed as a mirror of one's inner world---reflecting hidden, repressed feelings, attitudes, and refusals.

\subsection{Reflective Insight and Classification in the History of Science}

The author has penetrated through many of his own layers of belief systems through his many moments of self-reflection. Each change in his inner world (dissolution of beliefs or unconscious fear mechanisms) showed practically instantaneous change in his environment, an improvement in his sense of well-being, and an expansion of his ability to know.

For example, when someone pointed out his behavior and emphasized it with the statement ``I perceive it that way'', the author asked himself, ``What if I actually behave like this sometimes?'' This critical step of self-reflection---in contrast to earlier attempts to locate the cause of conflicts in others---led to a change in his behavior. Similarly, in discussions, the author tended to evaluate his own views as inherently more valid than those of others, which often led to strongly defending his arguments. By allowing himself the intention to understand the other person's perspective, he was able to integrate unfamiliar viewpoints, leading to more harmonious conversations that ended in mutual understanding rather than conflict.

This reflective introspection revealed to the author the nature of subjective perception: It is shaped by early experiences that filter reality as beliefs, while objective mechanisms---such as mirroring by others as a higher-order law---embed subjective experience in an interplay with reality.

This example not only shows the Reflective Method in action but also provides a philosophical argument for its relevance: Without the reflection of the subjective side, even empirical science remains blind to the totality of reality, which can only be understood through this interplay---an idea that expands the history of science by the role of the observer.

The introspective analysis of belief formation, as obtained in this example through Reflective Introspection, resonates with established theories and methods of behavioral psychology, which identify beliefs as central mechanisms of human behavior and experience, while simultaneously transcending their limits. Aaron T. Beck's Cognitive Behavioral Therapy (CBT) \cite{beck1976cognitive} postulates that dysfunctional beliefs---deeply rooted convictions about oneself, the world, and others---filter perception and trigger emotional reactions such as fear or depression. Similarly, Jeffrey E. Young's Schema Therapy \cite{young2007schema} describes early childhood schemas that arise from experiences such as parental resistance and shape behavior as cognitive patterns.

These approaches share with the author's analysis the emphasis on early childhood imprints, but they lack the precise recording of the dynamics of will and attention that the author uncovers through the reflection of the inner field of tension (``I want to, I cannot''). Furthermore, their perspective remains psychological and without the mathematical-philosophical depth that the author's work strives for.

Albert Ellis' Rational-Emotive Behavior Therapy (REBT) \cite{ellis2011rational} recognizes irrational beliefs as the cause of maladaptive behavior and aims at their reflection and modification---a goal that resembles the decision in the author's example to relativize one's own attitude. However, while REBT focuses on cognitive restructuring, the insights gained through the Reflective Method integrate somatic stress states. Acceptance and Commitment Therapy (ACT) (Hayes et al. \cite{hayes2012acceptance}), in turn, shares with the author's approach the acceptance of reality as a transformative force, but without the structural depth of the interactions between inner attitude and external mirroring that the author illuminates.

Sigmund Freud's psychoanalysis \cite{freud1915unconscious} touches on unconscious conflicts and defense mechanisms that appear as childhood decisions in the author's example, but it lacks the clear architecture of the ``I want'' field of tension that supports the reflections in this example. Somatic Experiencing \cite{levine1997waking}, finally, recognizes the role of the body in processing stress and trauma---an aspect that resonates in the infant's stress experience---but remains focused on healing, while the Reflective Method embeds this body experience in a more comprehensive understanding of consciousness, which captures the interaction between attention, will, and somatic stress through introspection.

It is noteworthy that these insights were obtained through the author's introspective introspection without prior knowledge of these theories, which underscores the originality of the Reflective Method. They go beyond this in several respects: For example, the precise representation of the will and attention dynamics (``I want to, I cannot'') models the formation of beliefs more systematically than CBT, by highlighting the tension between inner drive and external resistance as a formative principle. Likewise, the inclusion of physical analogies---such as the interactions in the ``multiplayer game'' of reality---opens up a dimension that is missing in ACT or psychoanalysis, as it understands the interaction between subjective experience and the objective world as an energetic system. The Reflective Method thus expands these approaches by using subjective reflection not only as a therapeutic tool but also as a source of knowledge that enriches scientific models and transcends their limits.

The formalized findings regarding the formation of beliefs, their anatomy, and the possibilities and mechanisms for dissolution will be presented in a further paper.

\subsection{A Pragmatic Approach to Recognizing One's Own Biases}

With the following method, or rather inner approach, one can recognize one's own biases and develop the ability of \textit{Reflective Empiricism}:

\begin{enumerate}
    \item \textbf{Encountering new information:} The researcher encounters information (e.g., facts, phenomena, theories, or explanatory models) that they reflexively reject. This rejection often occurs spontaneously and is supported by seemingly rational justifications.

    \item \textbf{Taking a step back:} Before fully focusing on the rejection, they consciously take a step back, recognizing that belief systems and other bias-generating mechanisms might influence their reaction.

    \item \textbf{Introspective self-questioning:} The researcher asks themselves the following questions to reflect on their rejection:
    \begin{itemize}
        \item \textit{Why do I reject this information?}
        \item \textit{Which of my assumptions or beliefs conflict with this information?}
        \item \textit{Are these assumptions fundamental axioms of my worldview, or are they well-considered, reflected, and evidence-based?}
        \item \textit{What specific evidence, facts, or research findings support my assumptions and justify my rejection?}
    \end{itemize}

    \item \textbf{Objective and reflective evaluation:} The researcher evaluates the new information both objectively and reflectively, considering its potential implications for their worldview:
    \begin{itemize}
        \item \textit{Is the source of this information credible?}
        \item \textit{Is my assessment of the source's credibility well-founded, or is it influenced by subjective biases?}
        \item \textit{What would it mean for me and my assumptions if this new information were actually true?}
    \end{itemize}

    \item \textbf{Conscious classification and decision-making:} Based on these introspective findings, the researcher builds a broader and more objective foundation for evaluating the new information. They then consciously decide what significance and value this information holds for them.
\end{enumerate}

Through this introspective approach, a responsible scientist---and indeed anyone in their daily life---can significantly enhance their capacity for insight and expand their horizon of understanding. By uncovering hidden patterns of thought and behavior, this method enables a deeper comprehension of both subjective experiences and their interplay with objective reality, fostering a more holistic and nuanced perspective of the world.

\section{Premise-Based Logical-Explorative Modeling}
\label{sec:modellbildung}

In order to concretize \textit{Reflective Empiricism} as a comprehensive methodology, it is necessary to illuminate the premise-based logical-explorative modeling presented here as a central methodological approach. This way of thinking represents an intentional and conscious deviation from established scientific methods and opens up an explorative space for the generation of new insights, especially in areas characterized by complex and elusive phenomena.

\subsection{Core Features of Premise-Based Logical-Explorative Modeling}

At the heart of this methodology is the explicit setting of premises as the starting point for knowledge acquisition. In contrast to the traditional empirical method, which is primarily based on external observational data and experimental findings, premise-based modeling begins with hypothetical assumptions or conceptual assumptions based on the observation of real phenomena. These premises are not necessarily empirically validated or directly verifiable, but serve in a first step as a framework for the exploration of logical consequences and the development of models.

The key features of this method can be summarized as follows:

\begin{itemize}
    \item \textbf{Premises as a Starting Point:} The knowledge process begins not primarily with empirical data, but with the conscious formulation of assumptions and premises grounded in observed phenomena. While these premises may extend into speculative or hypothetical territory, they are explicitly defined, logically coherent, and motivated by real-world observations.
    \item \textbf{Logical Reasoning as a Method:} Based on the premises set, logically consistent conclusions are derived and linked to a coherent model. The logical stringency and internal consistency of the model are paramount.
    \item \textbf{Exploration of the Conceptual Framework:} The method primarily aims at the exploration of a conceptual framework and the generation of new hypotheses and perspectives, not at immediate empirical validation or falsification. It serves to open up new possibilities of thinking and to enter unexplored terrain.
    \item \textbf{Model Building as a Form of Knowledge:} The result of the method is an explorative model that systematically presents the logical consequences of the premises and serves as a heuristic tool for further research. The model is preliminary and open to revision, but logically sound within the chosen conceptual framework.
    \item \textbf{Intuition as a Creative Source:} The setting of premises and the development of models can be inspired by intuitive insights and creative thought processes. The method recognizes the role of intuition as a legitimate source of scientific creativity, but at the same time emphasizes the need for logical stringency and systematic reflection.
\end{itemize}

\subsection{Differentiation from Established Scientific Methods}

Premise-based logical-explorative modeling differs significantly from established scientific methods and complements them in a specific way:

\begin{itemize}
    \item \textbf{Differentiation from the Empirical Method:} Unlike the empirical method, which is primarily data-driven and emphasizes objective observation and experimental verification, premise-based modeling starts with observed phenomena or established facts. To construct a coherent explanatory framework, it incorporates carefully chosen hypothetical auxiliary assumptions that logically extend these observations. These assumptions are not arbitrary but are chosen to logically and systematically explore possible explanations for phenomena that may not yet be fully understood or empirically verified. This makes the method particularly valuable in areas where direct empirical data is difficult to obtain or where current paradigms fall short of providing comprehensive explanations.
    \item \textbf{Differentiation from the Deductive Method:} Unlike the deductive method, which derives conclusions strictly from established axioms or principles, premise-based modeling incorporates hypothetical premises that are not treated as irrefutable truths but are rooted in observed phenomena and designed to explore their logical implications. These premises are introduced to create a logical structure capable of explaining complex or elusive phenomena. This approach enables a more flexible and explorative investigation of scenarios that lie beyond the reach of purely deductive reasoning.
    \item \textbf{Differentiation from the Inductive Method:} While the inductive method generalizes from specific observations to broader principles, premise-based modeling does not rely solely on observed data. Instead, it integrates observations with abstract or hypothetical premises that are designed to explore the logical implications of a given phenomenon. This hybrid approach retains a deductive structure but emphasizes exploration over generalization \cite{peirce1955philosophical}.
    \item \textbf{Differentiation from the Hypothetico-Deductive Method:} The hypothetico-deductive method formulates hypotheses based on empirical data and tests predictions derived from those hypotheses. In contrast, premise-based modeling begins with observed phenomena but allows for the introduction of auxiliary assumptions to construct an explanatory model. This model is developed logically and systematically before empirical testing becomes the focus. This sequence enables the exploration of phenomena that may resist immediate empirical validation while maintaining a clear methodological rigor.
\end{itemize}

\subsection{Value and Application of Premise-Based Logical-Explorative Modeling}

Premise-based logical-explorative modeling is not a substitute for established scientific methods, but a valuable addition and extension of the scientific toolkit. It is particularly suitable for:

\begin{itemize}
    \item Exploration of New Research Fields: In areas where little secure knowledge exists or new paradigms need to be developed, the method can help to open up new directions of thought and generate hypothetical models that can later be empirically investigated.
    \item Investigation of Complex and Elusive Phenomena: For phenomena that elude direct empirical observation or are methodologically difficult to access (such as consciousness, quantum phenomena, boundary areas of knowledge), the method can provide new perspectives and explanatory approaches that go beyond purely empirical approaches.
    \item Promotion of Scientific Creativity and Innovation: By consciously setting premises and exploring the conceptual framework, the method promotes creative and speculative thinking in science. It encourages researchers to develop bold and unorthodox ideas and to pursue new paths of knowledge acquisition.
\end{itemize}

Premise-based logical-explorative modeling is thus a valuable tool within \textit{Reflective Empiricism} to expand the boundaries of knowledge, open up new perspectives, and promote scientific creativity. It represents a methodological innovation that makes it possible to systematically and scientifically investigate even complex, speculative, and elusive phenomena.

\section{Connections to Existing Approaches: Classification in the History of Science}

\textit{Reflective Science} ties into historical and philosophical currents that already anticipated the limits of the objectivist paradigm:

\begin{itemize}
    \item Phenomenology: Husserl's Epoché \cite{husserl1960cartesian} emphasized introspection as access to experience, but the Reflective Method makes it the systematic foundation that unites subject and object---a step that Husserl did not fully exhaust philosophically.
    \item Quantum Physics: The observer effect \cite{stapp1993mind} showed that the role of the subject in physics is not negligible; \textit{Reflective Science} expands this into a holistic interplay that considers the development of quantum theory in a new context.
    \item Cognitive Psychology: Beck's belief theory \cite{neisser1967cognitive} describes subjective filters, but Reflective Reflection actively integrates them into knowledge acquisition, a further development that goes beyond psychological models.
\end{itemize}

The novelty lies in the systematic connection of these approaches to a paradigm that dissolves the illusion of pure objectivity and establishes intuition and paradigmatic openness as keys to universal insights---an approach that not only analyzes the history of science but actively continues it.

\section{The Transformative Potential of Reflective Science}

\textit{Reflective Science} opens doors to a comprehensive exploration of reality by overcoming the limits of a supposedly objective mind and using intuition and inner insights as a source---on the one hand, to recognize and reduce one's own bias filters and thereby arrive at a more objective understanding of superordinate mechanisms connected to the observer, and on the other hand, to gain deeper insights and a more comprehensive understanding of nature. A perspective that could also re-illuminate the history of physics as a sequence of intuitive breakthroughs.

It should be particularly emphasized that \textit{Reflective Science} does not aim to replace empirical research, but to complement it. By integrating subjective experiences and intuitive insights into the scientific process, the flexibility and speed of establishing new knowledge can be significantly increased. Paradigmatic blindness, as historically shown in the initial rejection of revolutionary theories, can be overcome through the systematic reflection of one's own perceptual filters and beliefs.

The delays that have arisen in the history of science due to resistance to new perspectives---be it with Galileo, Darwin, or Einstein---could be reduced by a scientific culture that integrates Reflective Methods. This culture would not regard anomalies as disruptive factors, but as potential signposts to new knowledge, and would not ignore the subjective dimension of research, but actively reflect and use it.

At a time when complex phenomena such as consciousness, global systems, or emergent properties are becoming the focus of science, the Reflective Method offers a way to overcome the limits of reductionist thinking and to develop more holistic explanatory approaches. It strives for a new balance between subjective depth and objective verifiability and thus represents an approach that meets the challenges of the 21st century.

\section{Case Studies in Premise-Based Modeling}
\label{sec:outcomes}

The method introduced in this paper did not remain theoretical. Over the following 14 months, systematic application of Reflective Empiricism and premise-based modeling yielded a coherent body of the author's own subsequent works, published across psychology, epistemology, ethics, and fundamental physics. The nine works surveyed below are presented not merely as results but as \textit{demonstrations of the method}: each case traces the phenomenological origin of the central insight, the premises derived from it, the logical exploration that followed, and the resulting model and its empirical relationship to established knowledge. The structural unity across such diverse domains---from cognitive psychology to special relativity---invites the question of whether a shared generative principle underlies such diverse results.

Each case study follows the same five-stage sequence that defines premise-based logical-explorative modeling: \textit{phenomenological constitution} (what observations made the premises accessible), \textit{premises and Eureka moment} (the central intuitive condensation), \textit{logical exploration} (the conceptual space that opens), \textit{model building} (formal elaboration), and \textit{validation} (relationship to known phenomena and empirical anchors).

\subsection{Psychology, Epistemology, and Ethics}

\subsubsection*{Belief Dynamics}

\textit{From Heuristic to Reflective Worldview: A Mathematical Model of Belief Dynamics}~\cite{wittwer2025belief-v2.1} traces the following generative path.

\textbf{Phenomenological constitution.} Years of rigorous self-observation of the author's own cognitive processes formed the foundation: How is new information processed? Why do some external inputs trigger immediate defensive devaluation while others are accepted? The author observed that such rejection tends to occur precisely when an incoming claim conflicts with a central existing core belief---and that the reaction is accompanied by specific emotional and somatic signatures. Sustained practice in suspending this reflex---deliberately asking ``what if this were true?''---sharpened the introspective resolution.

\textbf{Premises and Eureka moment.} The condensation occurred as a single image: if beliefs and information items are treated as vectors in an abstract space, the worldview can be understood as the totality of belief vectors, and agreement and contradiction as geometric alignment. A second immediate insight followed: representing agreement and contradiction via an inner product maps the relation onto the interval $[-1, 1]$, giving contradiction and confirmation a shared, graded measure, allowing them to be treated mathematically.

\textbf{Logical exploration.} From these two premises, the question became: how does a new information vector interact with the existing set of beliefs? Which subsets confirm, which contradict? How does the aggregate affect cognitive equilibrium? The model space opened naturally into dissonance as a function of two opposing sums, gatekeeper dynamics filtering information before full cognitive contact, and the structural attractors toward which belief systems drift.

\textbf{Model building.} Cognitive dissonance is operationalised as
\[
D = 4 \cdot C^{+} \cdot C^{-},
\]
maximal when confirming and contradicting beliefs pull with equal force. A gatekeeper mechanism
\[
W_I = W_{\text{source}} \cdot W_{\text{messenger}}
\]
formalises how perceived credibility filters information before cognitive processing begins. Iterative elaboration mapped belief islands, wall and bridge beliefs, and end-state attractors.

\textbf{Validation.} Twelve well-known phenomena---cognitive dissonance, confirmation bias, the backfire effect, motivated reasoning, compartmentalization, ad hoc rationalization, cognitive immunity and dogmatism, fragmentation and echo chambers, selective exposure, authority as a cognitive shortcut, immunization through source discrediting, and group polarization---emerge as structural consequences of the model, without being postulated. Beyond reproducing known phenomena, the model introduces six structural concepts not previously formalised: the gatekeeper mechanism as a precise bridge between sociological and individual filtering; belief islands with correlation-driven thinking dynamics; the wall/bridge belief distinction as the two archetypal dissonance-resolution strategies; numbed coexistence and genuine consensus as social end-state attractors; the qualitative distinction between heuristic and reflective immunity; and three quantitative worldview quality metrics ($Q_\text{frag}$, $Q_\text{refl}$, $Q_\text{resist}$).

\subsubsection*{Reflective Epistemology}

\textit{Reflective Epistemology: A Metatheoretical Model for Integrating Subjective and Objective Dimensions of Knowledge}~\cite{wittwer2025epistemology-v2.1} traces the following path.

\textbf{Phenomenological constitution.} The same practice of self-observation turned toward the act of thinking itself: How do thoughts arise? What selects among competing thoughts? What constructs a chain of reasoning? When is a thought considered true, partially true, or false---and what determines the threshold?

\textbf{Premises and Eureka moment.} The key insight was: what we hold as true is always preceded by a constitutive act---the \textit{decision to believe}---leading to the perception of what we believe as truth. This separates the epistemic dimension (what is held as true) from the ontic dimension (what is the case), and the two need not coincide. The insight drove a formal separation between layers of knowledge that most epistemological frameworks conflate.

\textbf{Logical exploration.} The consequences were explored: if belief is an act with a direction (intention) and a content (what is believed), then a mathematics of mind becomes possible---one that structures relations, modalities, and content qualities, and defines operators on them.

\textbf{Model building.} The result is a formal architecture featuring a proto-grammar of the mind (relations, modalities, content qualities) and three operators---attentional vector, intention vector, and belief field---governing a semantic thought-space $\mathbf{G_W}$. True objectivity is redefined not as elimination of subjectivity but as bias-reflected, intersubjectively validated coherence, operationalised through a four-dimensional C/T/B/R assessment matrix (Consistency, Truth of premises, Belief, Recognition).

\textbf{Validation.} Few established formal models exist at this level. The framework provides foundations for a ``mathematics of the mind'' that is compatible with phenomenological and analytic traditions while extending both.

\subsubsection*{Emergent Ethics}

\textit{Ontological Symmetry Breaking: The Emergence of Ethical Structures from Subjective I-Thou-It Projections}~\cite{wittwer2025ethics-v2.1} traces the following path.

\textbf{Phenomenological constitution.} The author observed that individuals and groups apply different ethical weights to different classes of entities---and that these differences are not merely cultural accidents but reflect underlying structural asymmetries in how subjectivity is attributed.

\textbf{Premises and Eureka moment.} The condensing image: consider all entities as structurally equivalent from an external vantage---a field of symmetry. The act of perceiving others from a subjective point of view (I) introduces different perceived qualities (I, Thou, It), \textit{breaking} this symmetry. 

\textbf{Logical exploration.} Different patterns of symmetry breaking yield different ethical configurations as necessary structural outcomes.

\textbf{Model building.} Formalisation in an I-Thou-It triad, extended and refined iteratively, yields six archetypal ethical systems---solipsistic, selective, conventional, empathic, self-sacrificing, and universal---as the complete set of structural configurations generated by differential subjectivity attribution. No external normative axioms are required.

\textbf{Validation.} The model aligns with sociological accounts of in-group/out-group dynamics and narcissism theory (psychology), and connects structurally to Hume's is-ought distinction (philosophy): ethics emerges from ontology rather than being imposed upon it, providing a constructive answer to the question Hume left open.

\subsection{Constructive Physics: The PRN Programme}
\label{sec:prn}

The four cognitive-science works above exemplify the method in domains with inherently qualitative phenomena. The PRN programme demonstrates that the same method generates \textit{quantitative} physical theory when applied to the structure of physical reality itself.

\textbf{Phenomenological constitution (shared across PRN).} Standard physics provides extraordinarily precise formalisms---but no account of what particles, space, time, forces, or fields \textit{are}. For the author, this constituted a genuine explanatory gap that resisted suppression. Space as a container was unsatisfying not as a physical complaint but as a thinking complaint: it maps a known experiential structure onto a formalism without explaining anything. The driving question became: are space, particles, and forces \textit{emergent phenomena} of some underlying process? Could space emerge without mechanistic gaps? Could quantum measurement be accounted for without ad hoc assumptions or unexplained gaps?

\textbf{Premises and Eureka moment (PRN core).} The condensation was an image of a relational network of process nodes. Any such node that reallocates itself (moves) must spend processing capacity on that reallocation---capacity unavailable for internal self-maintenance (proper time). This \textit{resource trade-off}, seen through the lens of the author's own experience of attention as a finite resource, immediately suggested that relativistic time dilation might be a consequence of a process budget rather than a geometric postulate. From this image, three functional channels were named: $I$ (internal process rate), $R$ (relocation rate), $C$ (environmental coupling). The hypothesis that their capacities are mutually constrained was explicit from the first formulation.

\textbf{Logical exploration.} If the three channels exhaust a normalised budget, what constraint form captures this? An initial linear form was proposed; it does not yield Lorentz symmetry. A quadratic form $I^2 + R^2 + C^2 = 1$ was examined---and does yield Lorentz kinematics in the $C=0$ limit. This was then subjected to two independent derivations to assess whether the quadratic form is an arbitrary choice or a structural necessity: (i) Chentsov's uniqueness theorem establishes that the Fisher information metric is the unique invariant Riemannian metric on the probability simplex; via the Hellinger embedding this yields the Euclidean $L^2$ norm on amplitudes---the budget constraint as an information-geometric necessity, independent of any lattice assumption. (ii) The unitarity condition $S^\dagger S = \mathbf{1}$ on a 6-port scattering node with $O_h$ symmetry independently generates the same quadratic form from group theory. The convergence of two structurally independent derivations on the same constraint is the strongest single epistemological argument in the programme.

\textbf{Model building.} The substrate is modelled as a discrete process-relational network (PRN) with unitary 6-port nodes arranged in a cubic lattice. The budget equation $I^2 + R^2 + C^2 = 1$ operates at each node. A variational dissonance functional establishes that the simple cubic lattice is the unique ground state of this architecture. From this single structure, the following results are derived constructively, without free continuous parameters:

\begin{itemize}[noitemsep, topsep=2pt]
    \item \textit{Special Relativity}~\cite{wittwer2026sr-v3.0}: time dilation ($d\tau/dt = I$), length contraction, invariance and numerical value of $c$, full Lorentz group, energy-momentum relation $E^2 = p^2c^2 + m^2c^4$. A structural feature absent from standard SR: the substrate defines an absolute rest frame (ontic asymmetry), yet this frame is experimentally undetectable because all instruments are budget-constrained on equal footing---a \textit{self-veiling mechanism} that resolves the longstanding tension between relativity and substrate models. To naturally resolve anisotropies in the cubic grid, the metallurgical grain model with cubic domains was introduced. 
    \item \textit{Lattice Uniqueness}~\cite{wittwer2026lattice-v1.0}: the cubic lattice $\mathbb{Z}^3$ is the unique global zero of the variational functional. Spatial dimension $d=3$ is \textit{derived} from two independent requirements: thermal stability and an integer topological-charge spectrum ($\pi_3(S^2) = \mathbb{Z}$). The budget equation re-emerges as a corollary of unitarity on this geometry---the constraint becomes a consequence of the substrate, not an assumption.
    \item \textit{Fine-Structure Constant}~\cite{wittwer2026alpha-v1.1}: $\alpha^{-1}$ emerges as a geometric fixed point from the intersection of two independent analytical strands---an impedance ansatz and a mean-field soliton equilibrium. The predicted value $\alpha^{-1} = 136.89$ deviates by $0.11\%$ from the CODATA measurement with no continuous fit parameters; the sole numerical input is the lattice site-percolation threshold $p_c = 0.3116$.
    \item \textit{Born's Rule}~\cite{wittwer2026born-v1.0}: $P(i|\psi\rangle) = |a_i|^2$ is reduced to a classical absorption theorem. Basin-mass processes on the probability simplex form bounded martingales almost surely absorbed at simplex vertices with absorption probability equal to the initial basin mass. Quantum measurement is not a primitive postulate but a derived consequence of network geometry.
    \item \textit{Classical Mechanics}~\cite{wittwer2026cm-v1.0}: Newton's three laws, the conservation laws, the Lagrangian and Hamiltonian formalisms, and Newtonian gravitation follow from the budget equation. Noether's theorem is applied in reverse: continuous symmetries of the macroscopic limit emerge from discrete $O_h$ symmetries of the lattice. Gravitation appears as the trace coupling of a scalar substrate field ($A_{1g}$ sector), suggesting it is a consequence of information geometry rather than a separate fundamental interaction.
\end{itemize}

\textbf{Validation and empirical anchor.} The programme reconstructs a broad range of established physical laws from a single constraint without independent postulates---a structural unification whose cross-disciplinary consistency is statistically highly improbable by chance. The critical empirical bridge is provided by \textit{Zeno-Amplified Sidereal Modulation}~\cite{wittwer2026zasm-v1.0}: the continuous quantum Zeno effect quadratically amplifies the kinematic slowdown from preferred-frame motion. The substrate predicts a \textit{sidereal} (not solar) modulation of quantum decay rates with amplitude $\Delta\Gamma/\Gamma \approx 3.6 \times 10^{-5}$ at Zeno factor $Z = 100$, accessible in tabletop quantum-optics experiments. Its characteristic $Z^2$ scaling distinguishes the signal from environmental noise. This prediction is absent from standard relativistic quantum field theory, making it a decisive discriminating test between the PRN substrate and the standard continuum framework.

\subsection{Work in Preparation}

Quantum mechanics (Paper~07) and electrodynamics (Paper~08) are currently in advanced development within the same PRN framework. Both derivation strands proceed from the resource constraint $I^2 + R^2 + C^2 = 1$ without additional postulates; core results are established and the manuscripts are being prepared for submission.

\section{Conclusion: A Call for Reflection}

The empirical methodology has precisely deciphered the material world, but its blind spots---the marginalization of subjective experience and the resistance to anomalies---limit its scope. Reflective thinking offers a transformative perspective by recognizing reality as an interplay of subjectivity and objectivity, encouraging science to reflect on its own methods and assumptions.

This approach calls on researchers to remain open to learning, to investigate unexplained phenomena, and to embrace self-reflection as a path to true objectivity. By rehabilitating the role of the observer and integrating intuition with scientific inquiry, \textit{Reflective Science} challenges the illusion of pure objectivity that has shaped empirical science for centuries. It identifies self-reflection and creativity as central to scientific discovery and invites a revolutionary rethinking of consciousness and reality.

At a time when science faces increasingly complex, interdisciplinary questions---from consciousness and quantum gravitation to global ecological systems---\textit{Reflective Science} provides an expanded toolkit. By complementing the strengths of the empirical method with the depth of subjective experience, it opens up new horizons of knowledge that have the potential to significantly shape and enrich the science of the 21st century.


\newpage

\begin{thebibliography}{31}
\providecommand{\natexlab}[1]{#1}
\providecommand{\url}[1]{\texttt{#1}}
\expandafter\ifx\csname urlstyle\endcsname\relax
  \providecommand{\doi}[1]{doi: #1}\else
  \providecommand{\doi}{doi: \begingroup \urlstyle{rm}\Url}\fi

\bibitem[Chalmers(1995)]{chalmers1995facing}
David~J. Chalmers.
\newblock Facing up to the problem of consciousness.
\newblock \emph{Journal of Consciousness Studies}, 2\penalty0 (3):\penalty0
  200--219, 1995.

\bibitem[Feyerabend(1975)]{feyerabend1975against}
Paul Feyerabend.
\newblock \emph{Against method: Outline of an Anarchistic Theory of Knowledge}.
\newblock Verso, 1975.

\bibitem[Omodeo and Tupikova(2016)]{omodeo2016cosmology}
Pietro~Daniel Omodeo and Irina Tupikova.
\newblock Cosmology and epistemology: A comparison between aristotle's and
  ptolemy's approaches to geocentrism.
\newblock In Matthias Schemmel, editor, \emph{Spatial Thinking and External
  Representation: Towards a Historical Epistemology of Space}, pages 145--174.
  Max-Planck-Gesellschaft zur Förderung der Wissenschaften, 2016.
\newblock \doi{10.34663/9783945561089-06}.

\bibitem[Vitruvius(1960)]{vitruvius1960ten}
Vitruvius.
\newblock \emph{The ten books on architecture}.
\newblock Harvard University Press, 1960.

\bibitem[Holton(1970)]{holton1973roots}
Gerald Holton.
\newblock The roots of complementarity.
\newblock \emph{Daedalus}, 99\penalty0 (4):\penalty0 1015--1055, 1970.

\bibitem[Chalmers(1996)]{chalmers1996conscious}
David~J. Chalmers.
\newblock \emph{The conscious mind: In search of a fundamental theory}.
\newblock Oxford University Press, 1996.

\bibitem[Kuhn(1962)]{kuhn1962structure}
Thomas~S. Kuhn.
\newblock \emph{The structure of scientific revolutions}.
\newblock University of Chicago Press, 1962.
\newblock \doi{10.7208/chicago/9780226458144.001.0001}.

\bibitem[Sharratt(1994)]{sharratt1994galileo}
Michael Sharratt.
\newblock \emph{Galileo: Decisive Innovator}.
\newblock Cambridge University Press, 1994.

\bibitem[Oreskes(1999)]{oreskes1999rejection}
Naomi Oreskes.
\newblock \emph{The Rejection of Continental Drift: Theory and Method in
  American Earth Science}.
\newblock Oxford University Press, 1999.
\newblock \doi{10.1093/oso/9780195117325.001.0001}.

\bibitem[Westfall(1980)]{westfall1980never}
Richard~S. Westfall.
\newblock \emph{Never at rest: A biography of Isaac Newton}.
\newblock Cambridge University Press, 1980.
\newblock \doi{10.1017/CBO9781107340664}.

\bibitem[Isaacson(2007)]{isaacson2007einstein}
Walter Isaacson.
\newblock \emph{Einstein: His life and universe}.
\newblock Simon and Schuster, 2007.

\bibitem[Polanyi(1958)]{polanyi1958personal}
Michael Polanyi.
\newblock \emph{Personal Knowledge: Towards a Post-Critical Philosophy}.
\newblock University of Chicago Press, Chicago, 1958.

\bibitem[Beck(1976)]{beck1976cognitive}
Aaron~T. Beck.
\newblock \emph{Cognitive therapy and the emotional disorders}.
\newblock International Universities Press, 1976.

\bibitem[Young et~al.(2007)Young, Klosko, and Weishaar]{young2007schema}
Jeffrey~E. Young, Janet~S. Klosko, and Marjorie~E. Weishaar.
\newblock \emph{Schema Therapy: A Practitioner's Guide}.
\newblock Guilford Press, 2007.

\bibitem[Ellis and Ellis(2011)]{ellis2011rational}
Albert Ellis and Debbie~Joffe Ellis.
\newblock \emph{Rational Emotive Behavior Therapy}.
\newblock American Psychological Association, 2011.

\bibitem[Hayes et~al.(2012)Hayes, Strosahl, and Wilson]{hayes2012acceptance}
Steven~C. Hayes, Kirk~D. Strosahl, and Kelly~G. Wilson.
\newblock \emph{Acceptance and Commitment Therapy: The Process and Practice of
  Mindful Change}.
\newblock Guilford Press, 2012.

\bibitem[Freud(1915)]{freud1915unconscious}
Sigmund Freud.
\newblock The unconscious.
\newblock \emph{SE}, 14:\penalty0 159--210, 1915.

\bibitem[Levine(1997)]{levine1997waking}
Peter~A. Levine.
\newblock \emph{Waking the Tiger: Healing Trauma}.
\newblock North Atlantic Books, 1997.

\bibitem[Peirce(1955)]{peirce1955philosophical}
Charles~Sanders Peirce.
\newblock \emph{Philosophical writings of Peirce}.
\newblock Dover Publications, 1955.

\bibitem[Husserl(1960)]{husserl1960cartesian}
Edmund Husserl.
\newblock \emph{Cartesian Meditations: An Introduction to Phenomenology}.
\newblock Martinus Nijhoff, 1960.
\newblock \doi{10.1007/978-94-017-4952-7}.

\bibitem[Stapp(1993)]{stapp1993mind}
Henry~P. Stapp.
\newblock \emph{Mind, matter, and quantum mechanics}.
\newblock Springer, 1993.
\newblock \doi{10.1007/978-3-662-08765-7}.

\bibitem[Neisser(1967)]{neisser1967cognitive}
Ulric Neisser.
\newblock \emph{Cognitive psychology}.
\newblock Appleton-Century-Crofts, 1967.
\newblock \doi{10.4324/9781315736174}.

\bibitem[Wittwer(2026{\natexlab{a}})]{wittwer2025belief-v2.1}
Oliver~Marc Wittwer.
\newblock From heuristic to reflective worldview: A mathematical model of
  belief dynamics.
\newblock \emph{Zenodo}, May 2026{\natexlab{a}}.
\newblock \doi{10.5281/zenodo.20327504}.
\newblock Preprint, v2.1, May 21, 2026.

\bibitem[Wittwer(2025{\natexlab{a}})]{wittwer2025epistemology-v2.1}
Oliver~Marc Wittwer.
\newblock Reflective epistemology: A metatheoretical model for integrating
  subjective and objective dimensions of knowledge.
\newblock \emph{Zenodo}, July 2025{\natexlab{a}}.
\newblock \doi{10.5281/zenodo.16269769}.
\newblock Preprint, v2.1, July 21, 2025.

\bibitem[Wittwer(2025{\natexlab{b}})]{wittwer2025ethics-v2.1}
Oliver~Marc Wittwer.
\newblock Ontological symmetry breaking: The emergence of ethical structures
  from subjective i-thou-it projections.
\newblock \emph{Zenodo}, July 2025{\natexlab{b}}.
\newblock \doi{10.5281/zenodo.16276132}.
\newblock Preprint, v2.1, July 21, 2025.

\bibitem[Wittwer(2026{\natexlab{b}})]{wittwer2026sr-v3.0}
Oliver~Marc Wittwer.
\newblock Constructive derivation of special relativity from resource
  constraints.
\newblock \emph{Zenodo}, March 2026{\natexlab{b}}.
\newblock \doi{10.5281/zenodo.18847405}.
\newblock Preprint, v3.0, March 3, 2026.

\bibitem[Wittwer(2026{\natexlab{c}})]{wittwer2026lattice-v1.0}
Oliver~Marc Wittwer.
\newblock The cubic lattice as unique global zero of a variational functional
  on six-port graphs.
\newblock \emph{Zenodo}, May 2026{\natexlab{c}}.
\newblock \doi{10.5281/zenodo.20322790}.
\newblock Preprint, v1.0, May 21, 2026.

\bibitem[Wittwer(2026{\natexlab{d}})]{wittwer2026alpha-v1.1}
Oliver~Marc Wittwer.
\newblock A self-consistent geometric fixed point near the inverse
  fine-structure constant from a 6-port substrate model.
\newblock \emph{Zenodo}, June 2026{\natexlab{d}}.
\newblock \doi{10.5281/zenodo.20559450}.
\newblock Preprint, v1.1, June 5, 2026.

\bibitem[Wittwer(2026{\natexlab{e}})]{wittwer2026born-v1.0}
Oliver~Marc Wittwer.
\newblock Born's rule as simplex-martingale absorption: A conditional
  derivation from reduced substrate dynamics.
\newblock \emph{Zenodo}, May 2026{\natexlab{e}}.
\newblock \doi{10.5281/zenodo.20324462}.
\newblock Preprint, v1.0, May 21, 2026.

\bibitem[Wittwer(2026{\natexlab{f}})]{wittwer2026cm-v1.0}
Oliver~Marc Wittwer.
\newblock Classical mechanics from a resource constraint: Newton's laws,
  gravitation, and the lagrangian formulation from the budget equation and a
  cubic-lattice substrate.
\newblock \emph{Zenodo}, May 2026{\natexlab{f}}.
\newblock \doi{10.5281/zenodo.20325179}.
\newblock Preprint, v1.0, May 21, 2026.

\bibitem[Wittwer(2026{\natexlab{g}})]{wittwer2026zasm-v1.0}
Oliver~Marc Wittwer.
\newblock Zeno-amplified sidereal modulation: A laboratory test for
  preferred-frame effects via quantum coherence monitoring.
\newblock \emph{Zenodo}, May 2026{\natexlab{g}}.
\newblock \doi{10.5281/zenodo.20321521}.
\newblock Preprint, v1.0, May 21, 2026.

\end{thebibliography}
\end{document}